# *Ab-initio* insights into the elastic, bonding, phonon, optoelectronic and thermophysical properties of SnTaS$_2$

M. I. Naher, M. Mahamudujjaman, A. Tasnim, R. S. Islam, S. H. Naqib*

Department of Physics, University of Rajshahi, Rajshahi 6205, Bangladesh

*Corresponding author email: salehnaqib@yahoo.com

**Abstract**

SnTaS$_2$ is a recently discovered layered semimetal exhibiting type-II low transition temperature superconductivity. Except some superconductivity related parameters, most of the physical properties, namely, elastic, mechanical, bonding, phonon dispersion, acoustic, thermophysical, and optical properties of SnTaS$_2$ are unexplored till now. In this study, we have investigated these hitherto unexplored properties of SnTaS$_2$ for the first time employing density functional theory (DFT) based first-principles method. SnTaS$_2$ is a mechanically stable, elastically anisotropic compound with strongly layered feature. The bond hardness and Vickers hardness have been calculated. The material under study is ductile, soft and highly machinable. The chemical bonding feature has mixed character with significant contribution coming from the ionic channel. Phonon dispersion curves disclose dynamical stability. Electronic band structure calculations show simple metallic character. The Fermi surface consists of both electron-like and hole-like sheets with varying degrees of dispersion. The low energy (including visible part of the spectrum) refractive index of SnTaS$_2$ is high. The reflectivity is fairly nonselective over a wide range of photon energy and the absorption coefficient is large in the mid ultraviolet region. The Debye temperature and thermal conductivity of SnTaS$_2$ are low. The electron-phonon coupling constant has been calculated. The compound under study possesses optical anisotropy with respect to the polarization direction of the incident electric field.

## 1. Introduction

In recent days, superconductors with nontrivial band structure have attracted considerable interest of condensed matter physics research community due to the possibility of realizing some novel quantum states, such as topological superconductors (TSC) and Majorana fermions [1,2]. It has been found that both topological insulators and topological semimetals possess nontrivial electronic band structure [1,3–5]. Based on the configuration of band crossing near the Fermi level topological semimetals can be classified into Dirac semimetals, Weyl semimetals, nodal line semimetals, etc. [6–11]. Besides novel electronic band structure features, such systems often possess a number physical properties (mechanical, thermophysical, optoelectronic) which are useful from the point of view of technological applications [12-15].

SnTaS$_2$ is a centrosymmetric layered superconductor with time-reversal symmetry. The presence of topological nodal lines and drumhead like surface states has been also observed in SnTaS$_2$ semimetal. SnTaS$_2$ is a type-II superconductor with superconducting transition temperature $T_c$ = 3.0 K [16]. However, it is instructive to note that the studies on the physical properties of SnTaS$_2$ are limited only to structural, electronic (band dispersion and electronic density of states), magnetization, electrical resistivity, specific heat measurements, and superconductivity [16-19]. There is a significant lack of understanding of most of the physical properties, namely, elastic, mechanical, bonding, phonon dispersion, acoustic, thermophysical, and optical properties of SnTaS$_2$. As far as potential applications are involved, a thorough understanding of elastic and mechanical properties including the anisotropy is important. Understanding of the bonding properties is useful to make sense of elastic constants and moduli. The Debye temperature and phonon spectra are related closely to superconductivity. The thermodynamic characteristics are essential since these lead one to understand the behavior of the solids at elevated temperature and pressure. In this work, a number of thermophysical properties, namely, Debye temperature, melting temperature, heat capacity, thermal expansion coefficient, Grüneisen parameter, dominant phonon mode, and minimum thermal conductivity (both isotropic and anisotropic) are investigated in detail. Besides, the energy dependent optical constants are studied to predict the material's response to incident photons. Further, bond population analysis has been carried out and the hardness of the compound has been calculated. None of these properties have been studied before. In this study, our main intention was to bridge this significant research gap and it constituted the primary motivation of the present work. This investigation has produced a large volume of novel results which can be very useful for the research community.

Most recently, we have studied some topological systems belonging to different classes and some interesting properties have been observed relevant to their possible applications [12-15]. It is interesting to note that the bulk physical properties studied so far did not depend on the details of spin-orbit coupling (SOC) in any significant way [12-15]. Inclusion of SOC reveals the nature of the surface electronic states and shows the topological aspects in the electronic band structure. These finer features do not affect the bulk physical properties considered herein to a large extent [12-16].

The rest of this manuscript has been divided into three sections. A brief description of density functional theory (DFT) based computational methods used in this investigation has been presented in Section 2. The results obtained are presented and analyzed in Section 3. Finally, in Section 4, the important findings of the present work are summarized and conclusions are drawn.

## 2. Computational method

Density functional theory (DFT) [20] is the most widely applied formalism for *ab-initio* calculations of various properties of crystalline solids. The ground state of crystalline system is found by solving the Kohn-Sham equation [21] with periodic boundary conditions (involving the Bloch states). In this study, the structural, elastic, electronic and optical properties of ternary SnTaS$_2$ have been explored by using DFT based CASTEP (CAmbridge Serial Total Energy Package) simulation code [22]. This code implements the total energy plane-wave pseudopotential method. We have used both local density approximation (LDA) and generalized gradient approximation (GGA) exchange-correlation functionals to check the suitability for geometry optimization. It is found that GGA provides better estimation of ground state structural parameters for SnTaS$_2$ than those obtained via LDA in comparison with experimental lattice constants and crystal cell volume. The most popular GGA functional known as the Perdew-Burke-Ernzerhof (PBE) scheme [23] tends to overestimate equilibrium volume whereas its modified version – the PBEsol [24] accounts for better equilibrium volume. In the PBEsol two of the parameters of PBE are changed in order to satisfy the constraints that are more congenial for solids. Except that, both functionals have the same analytical form [22]. Therefore, the results reported herein are obtained using the GGA-PBEsol scheme. Vanderbilt-type ultra-soft pseudopotential are used to represent electron-ion interaction [25]. This pseudopotential saves substantial computational time without compromising the accuracy of the calculations significantly [14,22].

To perform the pseudo atomic calculation the valence electron configurations of S, Sn and Ta atoms have been taken as [$3s^2 3p^4$], [$5s^2 5p^2$] and [$5d^3 6s^2$], respectively. The cut-off energy for plane wave expansion is taken up to 700 eV. For sampling the Brillouin zones of the compound under study 20×20×4 k-points mesh is used based on Monkhorst-Pack scheme [26]. The geometry optimization is carried out applying the Broyden-Fletcher-Goldferb-Shanno (BFGS) minimization technique [27], to get the lowest energy crystal structure of SnTaS$_2$. The structure is relaxed up to a convergence threshold of 5×10$^{-6}$ eV/atom for energy, 0.01 eV/Å for the maximum ionic Hellmann–Feynman force, 0.02 GPa for maximum stress and 5×10$^{-4}$ Å for maximum atomic displacement, with finite basis set corrections [28].

The independent single crystal elastic constants $C_{ij}$, bulk modulus $B$, and shear modulus $G$ are evaluated from the 'stress-strain' method embedded within the CASTEP code [29]. The electronic band structure, total density of states (TDOS) and partial density of states (PDOS) are calculated using the optimized geometry of SnTaS$_2$. The frequency dependent optical properties are extracted from the estimated complex dielectric function, $\varepsilon(\omega) = \varepsilon_1(\omega) + i\varepsilon_2(\omega)$, which describes the frequency/energy dependent interactions of photons with electrons. Using the Kramers-Kronig relationships, the real part $\varepsilon_1(\omega)$ of dielectric function $\varepsilon(\omega)$ is obtained from the imaginary part $\varepsilon_2(\omega)$. The imaginary part, $\varepsilon_2(\omega)$, is calculated from the momentum representation of matrix elements of transition between occupied and unoccupied electronic states by employing the CASTEP supported formula expressed as:

$$\varepsilon_2(\omega) = \frac{2e^2\pi}{\Omega\varepsilon_0} \sum_{k,v,c} |\langle \psi_k^c | \hat{u}.\vec{r} | \psi_k^v \rangle|^2 \, \delta(E_k^c - E_k^v - E) \qquad (1)$$

where, $\Omega$ is the volume of the unit cell, $\omega$ is frequency of the incident electromagnetic wave (photon), $e$ is the electronic charge, $\psi_k^c$ and $\psi_k^v$ are the conduction and valence band wave functions at a given wave-vector $k$, respectively. The conservation of energy and momentum during the optical transition is ensured by the delta function. When the dielectric function $\varepsilon(\omega)$ is known, all the other optical parameters such as refractive index, optical conductivity, reflectivity, absorption coefficient, and energy loss function can be computed from it. The Mulliken bond population analysis [30] has been employed widely to understand the bonding characteristics of solids. For SnTaS$_2$, we have used a projection of the plane-wave states onto a linear combination of atomic orbital (LCAO) basis sets [31,32]. The Mulliken bond population analysis can be performed with the help of Mulliken density operator written on the atomic (or quasi-atomic) basis:

$$P_{\mu\nu}^M(g) = \sum_{g'} \sum_{\nu'} P_{\mu\nu'}(g') S_{\nu'\nu}(g - g') = L^{-1} \sum_k e^{-ikg} (P_k S_k)_{\mu\nu'} \qquad (2)$$

and a charge designated on atom $A$ is defined as,

$$Q_A = Z_A - \sum_{\mu \in A} P_{\mu\mu}^M(0) \qquad (3)$$

where $Z_A$ is the charge of the nucleus.

## 3. Results and analysis

### *3.1 Structural Properties*

The crystal structure of SnTaS$_2$ is hexagonal with space group *P63/mmc* (No. 194). The unit cell of SnTaS$_2$ compound is shown in Figure 1. The Sn, Ta and S atoms are located at special sites of: 2a (0, 0, 0) and (0, 0, 1/2) for Sn, 2c ± (1/3, 2/3, 1/4) for Ta, and 4e ± (0, 0, z) and (0, 0, z+1/2) with z ≈ 0.33 for S. It is clear from the crystal structure that SnTaS$_2$ consists of alternative layers of Sn and S along *c*-axis. The cell contains two Sn atoms, two Ta atoms and four S atoms (two formula units). The estimated lattice parameters of SnTaS$_2$ with available experimental values [16,18] are presented in Table 1. The agreement with prior experimental results is quite good, lending support to the reliability of present calculations.

**Table 1**

Calculated and experimental lattice constants $a$, $b$, and $c$ all in Å, equilibrium volume $V_o$ (Å$^3$), total number of atoms in the cell, and bulk modulus $B$ (GPa) of SnTaS$_2$.

| Compound | $a$ | $b$ | $c$ | $V_o$ | No. of atoms | $B$ | Ref. |
|---|---|---|---|---|---|---|---|
| | 3.345 | 3.345 | 17.580 | 170.37 | 8 | 77.94 | This work |
| SnTaS$_2$ | 3.309 | 3.309 | 17.450 | - | - | - | [16]$^{\text{Expt.}}$ |
| | 3.307 | 3.307 | 17.442 | - | - | - | [18]$^{\text{Expt.}}$ |

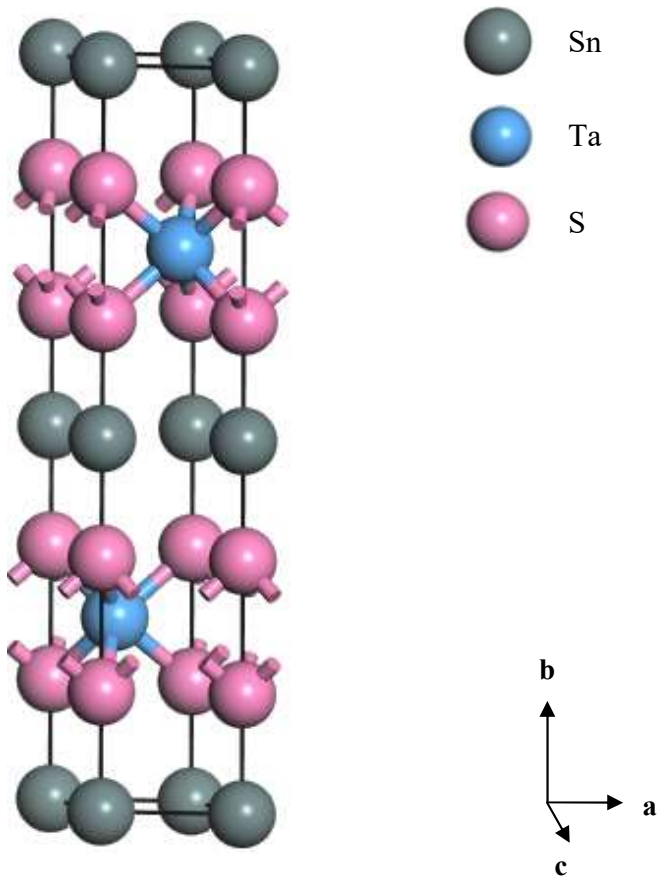

**Figure 1.** Three dimensional schematic crystal structure of SnTaS$_2$. The crystallographic directions are shown on the right.

## 3.2 Elastic Properties

In the determination of mechanical response and overall stiffness of a solid against externally applied forces elastic constants are of great importance. SnTaS$_2$ has a hexagonal crystal structure and from symmetry considerations there are five independent elastic constants ($C_{ij}$). These are: $C_{11}$, $C_{33}$, $C_{44}$, $C_{12}$, and $C_{13}$. The calculated elastic constants are listed in Table 2. For mechanical stability, according to Born-Huang conditions, a hexagonal system needs to satisfy the following inequality conditions [33]:

$$C_{11} - |C_{12}| > 0$$
$$(C_{11} + C_{12})C_{33} - 2C_{13}^2 > 0 \quad (4)$$
$$C_{44} > 0$$

All the elastic constants of SnTaS$_2$ have positive values and satisfy the stability criteria which indicate that SnTaS$_2$ is mechanically stable.

Among the five independent elastic constants, $C_{11}$ and $C_{33}$ measure the resistance of the material to the mechanical stress applied along the crystallographic *a*- and *c*-directions, respectively. For SnTaS$_2$, $C_{33}$ is greater than $C_{11}$ which indicates that the structure is more compressible along *a*-direction than that along the *c*-direction. Hence, the chemical bonding in the *c*-direction, i.e., the out-of-plane direction is considered to be greater than that in the *ab*-plane. In controlling the failure modes in solid such information is very useful. The resistance to shear deformation with respect to the tangential stress applied to the (100) plane in the [100] direction of the compound is indicated by the elastic constant $C_{44}$. It is seen that $C_{44}$ is lower than that of $C_{11}$ and $C_{33}$ and it indicates that the shear deformation takes place easily compared to unidirectional stress along any of the three crystallographic directions. The other elastic constants $C_{12}$, and $C_{13}$ are called off-diagonal shear components, which are related with compound's resistance due to various shape distortions.

**Table 2.** Calculated elastic constant, $C_{ij}$ (GPa), Cauchy pressure $C''$ (GPa), tetragonal shear modulus $C'$ (GPa) and Kleinman parameter ($\zeta$) for SnTaS$_2$.

| Compound | $C_{11}$ | $C_{33}$ | $C_{44}$ | $C_{12}$ | $C_{13}$ | $C'$ | $C''$ | $\zeta$ |
|---|---|---|---|---|---|---|---|---|
| SnTaS$_2$ | 153.65 | 202.83 | 11.57 | 43.24 | 26.76 | 55.21 | 31.67 | 0.43 |

The parameter $C'$ is a measure of crystal stiffness (the resistance to shear deformation by a shear stress applied in the (110) plane in the [1$\bar{1}$0] direction) and is given by,

$$C' = \frac{C_{11} - C_{12}}{2} \quad (5)$$

The brittleness/ductility of a compound can be understood by the Cauchy pressure, which is expressed as, $C'' = (C_{12} - C_{44})$. A positive Cauchy pressure corresponds to the ductility of the compound whereas the negative Cauchy pressure corresponds to the brittleness [34]. Calculated value of the Cauchy pressure for SnTaS$_2$ is positive hence the compound under investigation should be ductile in nature. The stability of a compound against stretching and bending can be measured by a parameter called the Kleinman parameter ($\zeta$) which is also known as internal strain parameter. The $\zeta$ of SnTaS$_2$ has been calculated using following equation [35]:

$$\zeta = \frac{C_{11} + 8C_{12}}{7C_{11} + 2C_{12}} \tag{6}$$

Kleinman parameter is a dimensionless parameter having values in the range of $0 \leq \zeta \leq 1$, where the insignificant contribution of bond bending to resist the external stress and the insignificant contribution of bond stretching to resist the external applied stress is represented by the lower and upper limits of $\zeta$, respectively.

The Hill approximated values of bulk modulus ($B_H$) and shear modulus ($G_H$) (using the Voigt-Reuss-Hill (VRH) method), Young's modulus ($Y$), and Poisson's ratio ($v$) of SnTaS$_2$ have been evaluated by the following standard formulae [36–38]:

$$B_H = \frac{B_V + B_R}{2} \tag{7}$$

$$G_H = \frac{G_V + G_R}{2} \tag{8}$$

$$Y = \frac{9BG}{(3B + G)} \tag{9}$$

$$v = \frac{(3B - 2G)}{2(3B + G)} \tag{10}$$

The resistance against tension or compression along its length of a material can be determined from Young's modulus. From Table 3, we can see that SnTaS$_2$ has neither a very high nor a very low value of Young's modulus [39-41]. Thus SnTaS$_2$ is capable of moderate resistance to large tensile stress.

The Poisson's ratio ($v$) plays a significant role in assessing mechanical properties of crystalline solids. The stability of solids against shear can be predicted from its value. The stability against shear is indicated by the low value of $v$. The nature of interatomic forces in solids is also related to Poisson's ratio [42,43]. In the case of solids which are dominated by central force, $v$ lies in the range from 0.25 to 0.50, whereas for non-central force solids, $v$ lies outside this range. Furthermore, in order to predict the failure mode of crystalline solids, the Poisson's ratio is an important tool. The materials whose Poisson's ratio is less than the critical value of 0.26 are expected to undergo brittle failure whereas those with Poisson's ratio greater than the critical value undergo ductile failure. The calculated value of $v$ of the compound $SnTaS_2$ is 0.31, which predicts the ductile failure of the compound and indicates that central force dominates in the atomic bondings of $SnTaS_2$. In purely covalent crystals, the Poisson's ratio is around 0.10 and in completely metallic compounds; the Poisson's ratio is around 0.33. Since the compound under study has $v = 0.31$, there is significant contribution of metallic bonding in $SnTaS_2$. The Pugh's ratio is defined by the simple relationship between bulk modulus and shear modulus given by $B/G$ [44–46]. The ductility or the brittleness of a solid can be determined from this particular ratio. The critical value of the Pugh's ratio is 1.75. Materials having a value higher than this show ductility, whereas materials having a value lower than this exhibit brittleness. From Table 3, we can see that for $SnTaS_2$ the Pugh's ratio is 2.37, which is substantially higher than the critical value. Thus $SnTaS_2$ is ductile in nature. High level of ductility of $SnTaS_2$ stems from ionic and metallic bondings in this compound.

The ease at which a material can be machined using cutting/shaping tools is quantified by a parameter known as the machinability index, $\mu_M$. As machinability index defines the optimum economic level of machine utilization, cutting forces, temperature and power strain, the information regarding machinability of a material is becoming valuable in today's industry. The machinability of a material is expressed in terms of machinability index [47]:

$$\mu_M = \frac{B}{C_{44}} \tag{11}$$

The machinability index of $SnTaS_2$ is 6.74. Compared to many other metallic ternaries, the machinability index of $SnTaS_2$ is very high [48-51]. It implies that this compound has high potential to be used as a dry lubricant. We have summarized the elastic moduli, ratios and machinability index in Table 3.

**Table 3.** The calculated values of bulk modulus ($B$) and shear modulus ($G$) (by Voigt-Reuss-Hill (VRH) method), Pugh's ratio ($B/G$), Young's modulus ($Y$), Poisson's ratio ($v$) and machinability index ($\mu_M$) of $SnTaS_2$.

| compound | $B_R$ | $B_V$ | $B_H$ | $G_R$ | $G_V$ | $G_H$ | $B/G$ | $Y$ | $v$ | $\mu_M$ |
|---|---|---|---|---|---|---|---|---|---|---|
| $SnTaS_2$ | 77.704 | 78.184 | 77.944 | 22.593 | 43.227 | 32.910 | 2.368 | 86.548 | 0.315 | 6.74 |

## 3.3 Elastic anisotropy

In the case of crystalline solids, information on elastic anisotropy is an important issue. The directional dependence of physical properties of a solid can be explained with the help the anisotropy parameters. It is well known that the anisotropic elastic properties of a crystal are related to a number of physical properties such as behavior of micro-cracks in crystals, propagation of cracks, development of plastic deformation in crystals etc. The shear anisotropic factors measure the degree of anisotropy in the bonding between atoms in different crystal planes. The shear anisotropy for a hexagonal crystal can be qualified by three different factors [52,53]. The shear anisotropy factor for {100} shear planes between the [011] and [010] directions is,

$$A_1 = \frac{4C_{44}}{C_{11} + C_{33} - 2C_{13}} \tag{12}$$

The shear anisotropy factor for {010} shear planes between the [101] and [001] directions is,

$$A_2 = \frac{4C_{55}}{C_{22} + C_{33} - 2C_{23}} \tag{13}$$

The shear anisotropy factor for the {001} shear planes between [110] and [010] directions is,

$$A_3 = \frac{4C_{66}}{C_{11} + C_{22} - 2C_{12}} \tag{14}$$

The calculated shear anisotropy factors ($A_i$) of SnTaS$_2$ are recorded in Table 4. The crystals having isotropy in the bondings existing between the planes have unit values ($A_1 = A_2 = A_3 = 1$) and any other value (lesser than or greater than unity) implies that there is anisotropy possessed by the crystal. The calculated values of $A_1$, $A_2$ as shown in Table 4 indicate that the compound under investigation is highly anisotropic. It is interesting to note that, $A_3$ is 1. The universal log-Euclidean index is given by the following equation [54,55]:

$$A^L = \sqrt{\left[\ln\left(\frac{B^V}{B^R}\right)\right]^2 + 5\left[\ln\left(\frac{C_{44}^V}{C_{44}^R}\right)\right]^2} \tag{15}$$

where, the Voigt and Reuss approximated values of $C_{44}$ are obtained from [54]:

$$C_{44}^R = \frac{5}{3}\left\{\frac{C_{44}(C_{11} - C_{12})}{3(C_{11} - C_{12}) + 4C_{44}}\right\} \tag{16}$$

and

$$C_{44}^V = C_{44}^R + \frac{3}{5}\left\{\frac{(C_{11} - C_{12} - 2C_{44})^2}{3(C_{11} - C_{12}) + 4C_{44}}\right\} \tag{17}$$

The expression for $A^L$ is valid for all crystal symmetries. The universal anisotropy index, $A^U$ is closely related to $A^L$. In the case of perfectly anisotropic crystal, $A^L = 0$. The values of $A^L$ ranges between 0 and 10.26, and almost 90% of crystalline solids have $A^L < 1$. It is also argued that, $A^L$ indicates whether the compound is layered/lamellar type [54,56,57]. The compounds that have higher value of $A^L$ indicate that they are strongly layered whereas the compounds that have low value of $A^L$ indicate that they have non-layered structural features. Since the calculated value of $A^L$ as shown in Table 4 is significantly larger than unity, we thus conclude that SnTaS$_2$ will exhibit layer type of structural properties.

The universal anisotropy index $A^U$, anisotropy in compressibility $A^B$, anisotropy in shear $A^G$ (or $A^C$) and equivalent Zener anisotropy measure $A^{eq}$ for material with any crystal symmetry can be estimated using following widely employed standard formulae [54,58,59]:

$$A^U = 5\frac{G_V}{G_R} + \frac{B_V}{B_R} - 6 \geq 0 \tag{18}$$

$$A^B = \frac{B_V - B_R}{B_V + B_R} \tag{19}$$

$$A^G = \frac{G^V - G^R}{2G^H} \tag{20}$$

$$A^{eq} = \left(1 + \frac{5}{12}A^U\right) + \sqrt{\left(1 + \frac{5}{12}A^U\right)^2 - 1} \tag{21}$$

**Table 4.** Shear anisotropy factors ($A_1$, $A_2$ and $A_3$), the universal anisotropy index $A^U$, anisotropy in compressibility $A_B$, anisotropy in shear $A_G$ (or $A^C$), equivalent Zener anisotropy measure $A^{eq}$ and universal log-Euclidean index $A^L$ for SnTaS$_2$.

| compound | $A_1$ | $A_2$ | $A_3$ | $A^U$ | $A^B$ | $A^G$ | $A^{eq}$ | $A^L$ |
|---|---|---|---|---|---|---|---|---|
| SnTaS$_2$ | 0.1527 | 0.1527 | 1 | 4.5726 | 0.0031 | 0.3134 | 5.6329 | 2.5636 |

The zero value of $A_B$ and $A_G$ indicates elastic isotropy, whereas a value of 1 indicates highest possible anisotropy. Therefore, from the values of $A_B$ and $A_G$ listed in Table 4 we can see that the compound under study is anisotropic in compressibility and shear. An anisotropy factor $A^U$, known as the universal anisotropy factor because of its applicability to all kinds of crystal symmetries was introduced by Ranganathan and Ostoja-Starzewski [58]. $A^U = 0$ implies elastic isotropy while any other positive value indicates elastic anisotropy. Similarly, zero value of $A^{eq}$ indicates elastic isotropy while any other positive value indicates its anisotropic nature. The calculated values of both $A^U$ and $A^{eq}$ indicate that the compound under investigation is anisotropic in nature.

The uniaxial bulk modulus along $a$, $b$ and $c$ axes of a solid and anisotropies in the bulk modulus are calculated by the following equations [53]:

$$B_a = a\frac{dP}{da} = \frac{\Lambda}{1+\alpha+\beta} \quad ; \quad B_b = a\frac{dP}{db} = \frac{B_a}{\alpha} \quad ; \quad B_c = c\frac{dP}{dc} = \frac{B_a}{\beta} \tag{22}$$

and

$$B_{relax} = \frac{\Lambda}{(1+\alpha+\beta)^2} \tag{23}$$

where, $\Lambda = C_{11} + 2C_{12}\alpha + C_{22}\alpha^2 + 2C_{13}\beta + C_{33}\beta^2 + 2C_{33}\alpha\beta$

$$\alpha = \frac{(C_{11}-C_{12})(C_{33}-C_{13}) - (C_{23}-C_{13})(C_{11}-C_{13})}{(C_{33}-C_{13})(C_{22}-C_{12}) - (C_{13}-C_{23})(C_{12}-C_{23})}$$

and

$$\beta = \frac{(C_{22}-C_{12})(C_{11}-C_{13}) - (C_{11}-C_{12})(C_{23}-C_{12})}{(C_{22}-C_{12})(C_{33}-C_{13}) - (C_{12}-C_{23})(C_{13}-C_{23})}$$

From the calculated values of $B_a$, $B_b$ and $B_c$, it is seen that $B_c$ is greater than both $B_a$ and $B_b$ which suggests that the compound SnTaS$_2$ is more compressible when stress is applied along $a$- and $b$-directions than along $c$-direction. The parameters $\alpha$ and $\beta$ are the measures of relative change of the $b$- and $c$- axis as a function of the deformation of the $a$-axis.

The anisotropies of the bulk modulus along $a$- and $c$-axis are defined as [53]:

$$A_{B_a} = \frac{B_a}{B_b} = \alpha \tag{24}$$

$$A_{B_c} = \frac{B_c}{B_b} = \frac{\alpha}{\beta} \tag{25}$$

Here, $A_{B_a}$ and $A_{B_c}$ represent the anisotropies of bulk modulus along the $a$-axis and the $c$-axis with respect to the $b$-axis, respectively. The linear compressibility of the compound along $a$- and $c$-axes ($\beta_a$ and $\beta_c$) are calculated from [60]:

$$\beta_a = \frac{C_{33} - C_{13}}{D} \quad \text{and} \quad \beta_c = \frac{C_{11} + C_{12} - 2C_{13}}{D} \tag{26}$$

with $D = (C_{11} + C_{12})C_{33} - 2(C_{13})^2$

From the calculated values as shown in Table 5, we can see that for the compound under investigation, the compressibility along *a*-axis is greater than that along *c*-axis.

**Table 5.** The bulk modulus ($B_{relax}$ in GPa), bulk modulus along *a*-, *b*- and *c*-axis ($B_a$, $B_b$, $B_c$ in GPa), anisotropy in bulk modulus along *a*- and *c*- axes ($A_{B_a}$ and $A_{B_c}$ in GPa), and linear compressibility along *a*- and *c*- axes ($\beta_a$ and $\beta_c$ in TPa$^{-1}$) for the compound SnTaS$_2$.

| Compound | $B_a$ | $B_b$ | $B_c$ | $B_{relax}$ | $A_{B_a}$ | $A_{B_c}$ | $\beta_a$ | $\beta_c$ |
|---|---|---|---|---|---|---|---|---|
| SnTaS$_2$ | 218.676 | 218.676 | 268.578 | 77.705 | 1 | 1.228 | 4.573 | 3.724 |

For isotropic solids, three dimensional (3D) direction dependent Young's modulus, shear modulus, compressibility (inverse of bulk modulus) and Poisson's ratio should exhibit spherical shapes, while any deviation from spherical shape would indicate anisotropy.

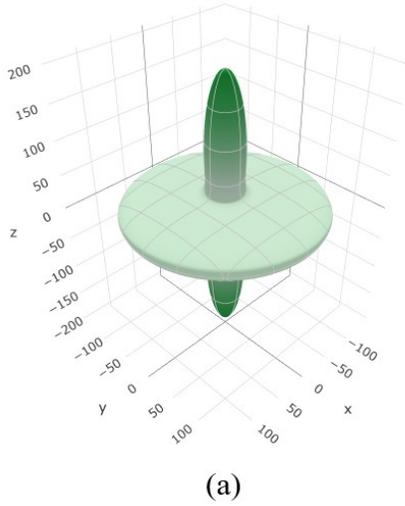

(a)

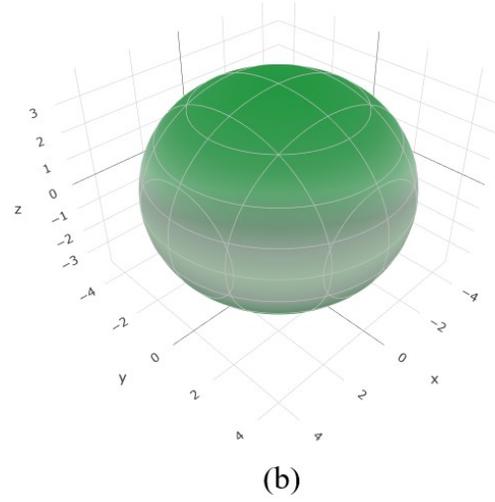

(b)

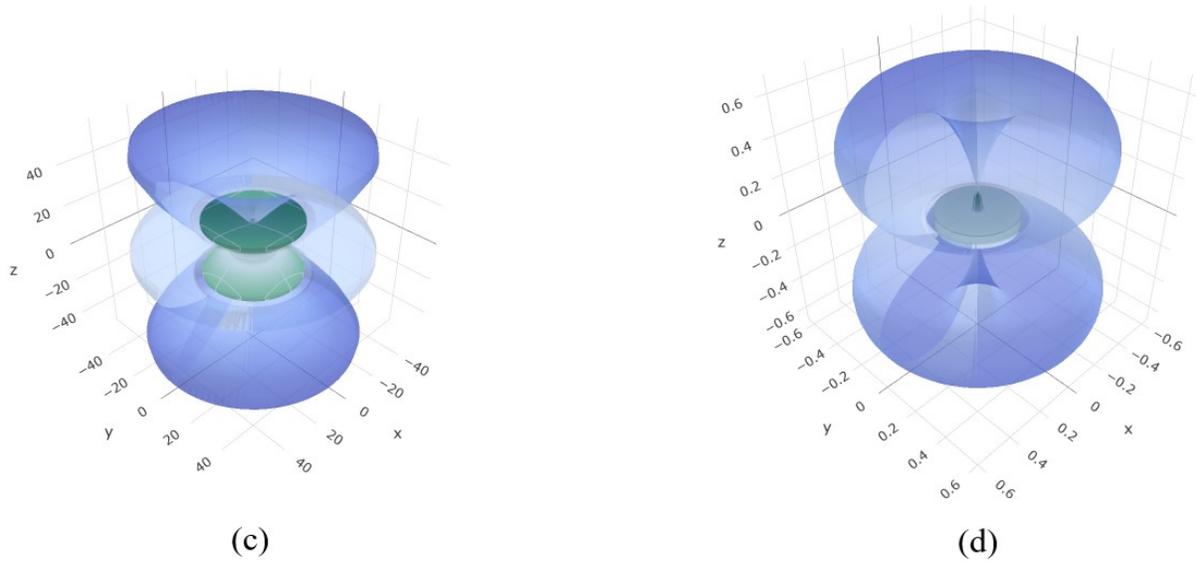

(c)  (d)

**Figure 2.** Three dimensional (3D) directional dependences of (a) Young's modulus (b) compressibility (c) shear modulus and (d) Poisson's ratio for $SnTaS_2$ compound.

We have shown ELATE [61] generated 3D plots of directional dependence of Young's modulus, shear modulus, compressibility and Poisson's ratio for $SnTaS_2$ in Fig. 2 above. Compared to other measures, 3D plot of Young's modulus exhibits very high level of anisotropy. This is originating from the layered character of the structure of $SnTaS_2$. The anisotropy in compressibility is the least, whereas Poisson's ratio and shear modulus possess significant out-of-*ab*-plane anisotropy.

## *3.4 Electronic properties*

### *3.2.1 Band structure*

Many important physical features, such as optical properties and charge transport properties of solids can be understood from the details of electronic band structure. Figure 3 shows the band structure diagram of $SnTaS_2$ along different high symmetry directions in the first Brillouin zone (BZ) for the optimized crystal structure in the ground state. The horizontal line drawn at 0 eV is the Fermi level, $E_F$. The band structure of the compound under study clearly manifests that several conduction band and valence band (band curves other than blue) cross the Fermi level. Therefore, $SnTaS_2$ should exhibit metallic character. The energy bands around the Fermi level mainly come from Ta-$5d$ states. Highly dispersive nature of the bands crossing the Fermi level near $\Gamma$-point for $SnTaS_2$ is indicative of high mobility of the charge carriers in this compound. It is interesting to note that the band crossing near the $\Gamma$-point displays hole-like feature. The bands in the conduction bands (CBs) running within *ab*-plane of the crystal (*A-H*, *K-$\Gamma$-M*) are highly dispersive. On the other hand, the band curves along *c*-direction ($\Gamma$-*A*, *H-K*, and *M-L*) are weakly dispersive in the CBs, which indicates very high effective mass of charge carriers and low mobility in this direction. Therefore, in addition to significant elastic/mechanical anisotropy, $SnTaS_2$ is expected to show anisotropy with respect to charge transport within and out of the *ab*-plane.

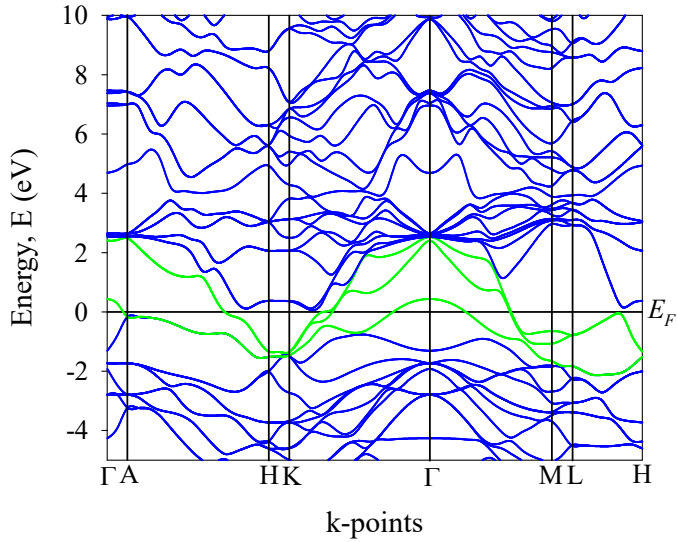

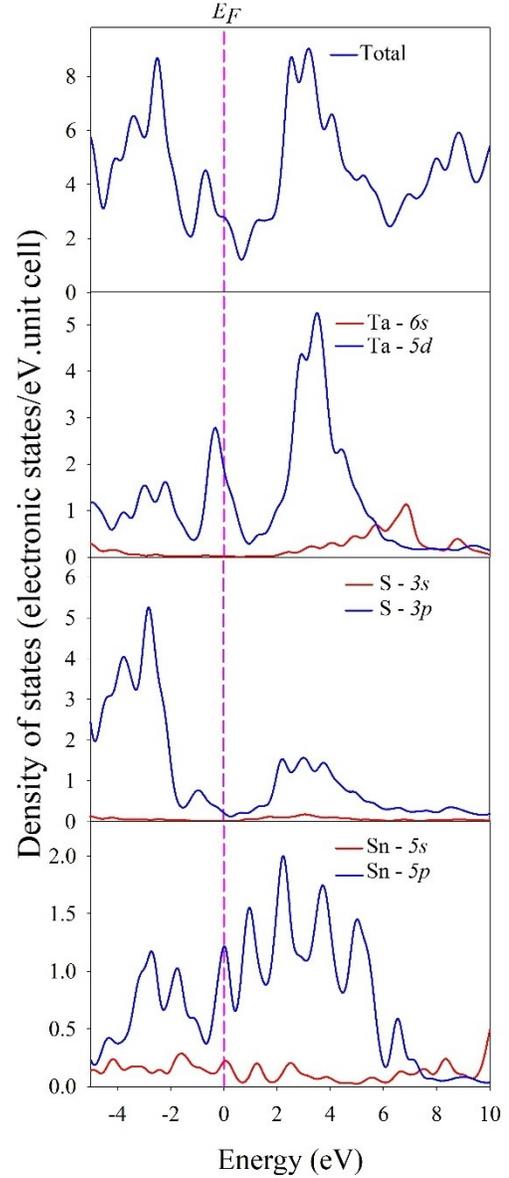

**Figure 3.** The band structure of $SnTaS_2$ along the high symmetry directions of the k-space within the first Brillouin zone.

**Figure 4.** Total and partial electronic density of states of $SnTaS_2$.

## 3.2.2 Density of states (DOS)

The calculated total density of states (TDOS) and atom resolved partial density of states (PDOS) of SnTaS$_2$ is shown in Figure 4. The vertical broken line drawn at 0 eV represents the Fermi level, $E_F$. The finite value of TDOS at the Fermi level indicates SnTaS$_2$ is a metallic compound. The TDOS value of the compound at $E_F$ is found to be 2.8 states per eV per unit cell or 1.4 states per eV per formula unit; in very good agreement to that calculated in Ref. [16] both theoretically and experimentally. This low value of $N(E_F)$ comes from the highly dispersive bands crossing the Fermi level. The main contributions to the TDOS stem from S-*3p*, Ta-*5d*, and Sn-*5p* states in the formation of valence band, while the conduction band mainly consists also of Ta-*5d*, S-*3p* and Sn-*5p* electronic states. There is a significant hybridization between these states. The gap between the bonding state and anti-bonding state is known as pseudogap. The electronic stability of a solid is related to the presence of a pseudogap or quasi-gap in the TDOS around the Fermi level [62,63]. In SnTaS$_2$ bonding and anti-bonding peaks are within ~1.5 eV from the Fermi level. The close proximity of the peaks in the TDOS to the Fermi energy indicates that it might be possible to change the electronic phase of this compound by suitable atomic substitution (alloying) or by applying pressure. The electron–electron Coulomb interaction parameter can be calculated using the following relation [64] :

$$\mu^* = \frac{0.26 N(E_F)}{1 + N(E_F)} \qquad (27)$$

The electron-electron interaction parameter, also known as the Coulomb pseudopotential, of SnTaS$_2$ is found to be 0.15. The computed value of $\mu^*$ is slightly high compared to those of many other compounds exhibiting conventional superconductivity at low temperatures [65]. This repulsive Coulomb pseudopotential reduces the transition temperature, $T_c$, of superconducting compounds [50, 65-67].

## 3.2.3 Fermi surface

Fermi surface topology of a material dominates large number of electronic, transport, optical and even magnetic properties. We have also constructed the Fermi surface of SnTaS$_2$ from the electronic band structure, as shown in Figure 5. Fermi surfaces are constructed from band number 20, 21 and 22 of SnTaS$_2$ which cross the Fermi level. It is seen that SnTaS$_2$ contains both electron- and hole-like Fermi sheets. The central sheet exhibits electron-like feature. However, the external sheet with hexagonal symmetry mainly shows hole-like character. There is a notable anisotropy in the Fermi surface topology.

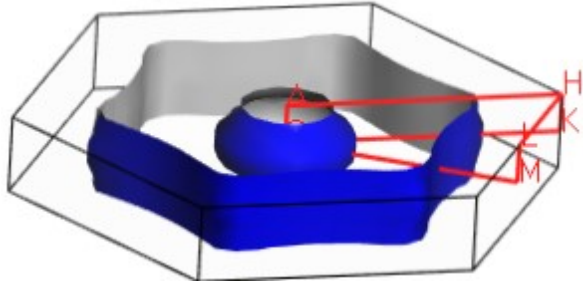

**Figure 5.** The Fermi surface of SnTaS$_2$.

## *3.2.4 Electronic charge density distribution*

The charge density distribution illustrates the nature of bonding among different atoms. We have calculated charge density to explain the transfer of charge among atoms and to visualize the bonding characteristics of SnTaS$_2$. The electronic charge density distribution of SnTaS$_2$ is shown in Figure 6. The color scale on the right hand side of charge density maps discloses the total electron density (blue and red color indicate high and low charge (electron) density, respectively). The accumulation of charges between two atoms constitutes covalent bonds, whereas the balancing of positive or negative (depletion regions) charge at the atomic position indicates ionic bonding. Weak indication of covalent bonding is found between S-S, S-Ta and Sn-S atoms. Accumulation/depletion of charge around S and Sn atoms is indicative of strong ionic bonding between these atoms. Maximum electron density is observed around S atoms as compared to other atoms. The overall features are consistent with the DOS curves (Fig. 4), which shows hybridization between different atomic orbitals of Sn, Ta, and S in the valence band of SnTaS$_2$.

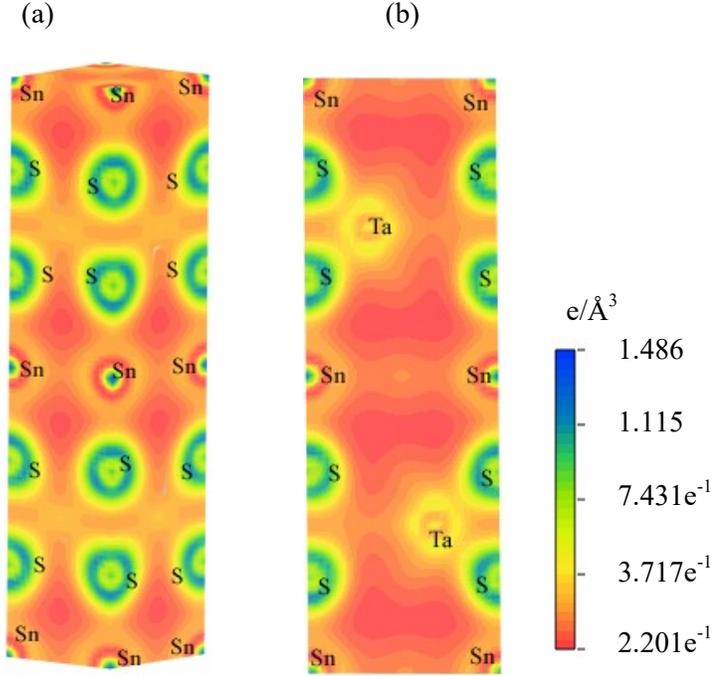

**Figure 6.** The electronic charge density distribution map (e/Å$^3$) of SnTaS$_2$ in (a) three dimensional visualization and (b) (111) plane.

## *3.5 Phonon dispersion curves and phonon density of states*

The characteristics of phonons have fundamental importance in the study of crystalline materials. The electron–phonon interaction function is directly related to the phonon density of states (DOS). Various physical properties of materials can be determined directly or indirectly from the phonon dispersion spectra and phonon DOS [68]. With the help of phonon dispersion spectra (PDS) the structural stability, phase transition and contribution of vibrations in thermal and charge transport properties of a material can be explained [69]. The phonon dispersion curves and phonon density of state (total and partial) of SnTaS$_2$ in the ground state have been calculated applying the density functional perturbation theory (DFPT) based finite displacement method (FDM) [70,71]. The PDC along with the total phonon density of states (PHDOS) of SnTaS$_2$ along the high symmetry directions of the Brillouin zone (BZ) at zero pressure is illustrated in Figure 7. We have put phonon dispersion curve and phonon density of states graph side by side in order to compare the bands and their corresponding density of states. Generally, the existence of negative frequency at gamma point in PDS indicates that the compound is dynamically unstable. Since the PDS of SnTaS$_2$ exhibit positive phonon frequencies in the whole region of the Brillouin zone, SnTaS$_2$ is a dynamically stable compound. There is a clear spectral gap between the acoustic and optical phonon branches. The high frequency optical modes originate mainly from the vibrations of the S atoms whereas the acoustic branches are dominated by the Ta and Sn atoms as seen from the PHDOS profiles.

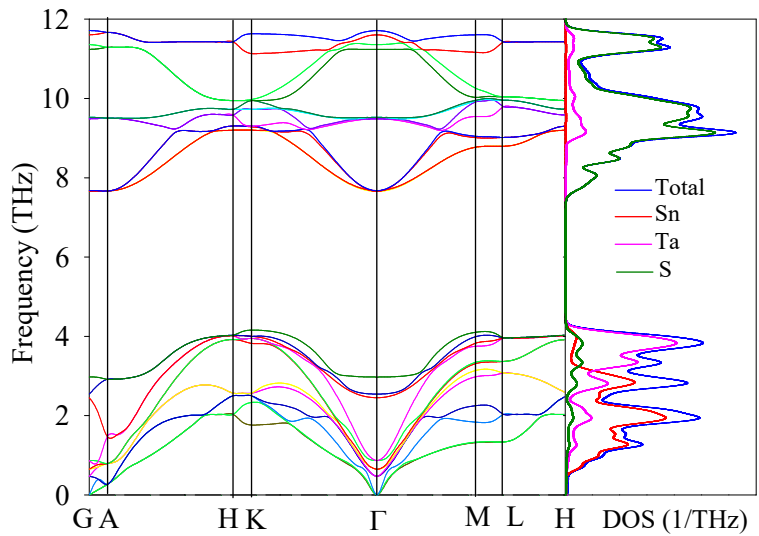

**Figure 7.** Calculated phonon dispersion spectra and phonon DOS of SnTaS$_2$ in the ground state.

## 3.6 Bond population analysis

We have used Mulliken population analysis (MPA) [30] to get understanding of the bonding nature (ionic, covalent and metallic) in SnTaS$_2$ and have calculated the effective valence of atoms in the molecule of interest. The results are shown in Table 6. It is worth mentioning that the charge spilling parameter represents the quantity of missing valence charges in a projection in the MPA. The lower value of charge spilling parameter indicates a good representation of electronic bonds. The charge spilling parameter for SnTaS$_2$ is only 0.59%.

**Table 6.** Charge spilling parameter (%), orbital charges (electron), atomic Mulliken charge (electron), Hirshfeld charge (electron), and effective valence charge (EVC) (electron) of SnTaS$_2$.

| Compound | Species | Charge spilling | s | p | d | Total | Mulliken charge | EVC | Hirshfeld charge |
|---|---|---|---|---|---|---|---|---|---|
|  | S |  | 1.82 | 4.35 | 0.0 | 6.17 | -0.17 | +1.83 | -0.11 |
| SnTaS$_2$ | Sn | 0.59 | 1.69 | 2.02 | 0.0 | 3.71 | 0.29 | +3.71 | 0.10 |
|  | Ta |  | 0.50 | 0.54 | 3.91 | 4.95 | 0.05 | +3.95 | 0.12 |

Atomic charges for S, Sn and Ta in the compound are -0.17, 0.29 and 0.05 respectively. Here, Sn and Ta atoms transfer electronic charge to S atom. This is an indication of partial presence of ionic bonds between different atoms of the compound. These charges come from *5p* orbital of Sn and *5d* orbital of Ta. The Mulliken atomic charges for S, Sn and Ta are deviated from their purely ionic state (S: -2, Sn: +4 and Ta: +4). This suggests that some covalent bonds are also present among S, Sn and Ta atoms. The effective valence charge (EVC) of SnTaS$_2$ is calculated to perceive the degree of covalency and/or ionicity. The difference between formal ionic charge and calculated Mulliken charge is called the EVC [30,32]. Zero EVC means a perfect ionic bond while non-zero value of EVC implies increasing level of covalency. From the above discussion it is clear that, covalent bonds are present in the compound under study. However, due to the strong basis set dependency, Mulliken population analysis sometimes gives results in contradiction to chemical intuition and overestimates covalency to a significant extent.

For these reasons, we have also employed Hirshfeld population analysis (HPA) [72] because it has practically no basis set dependence and often gives more meaningful results. According to HPA, atomic charges for S, Sn and Ta in SnTaS$_2$ are -0.11, 0.10 and 0.12, respectively. These values are significantly lower than those predicted by the MPA. However, Hirshfeld charge also predicts that electrons are transferred from Sn and Ta to S atom. This result is consistent with that of MPA.

## 3.7 Theoretical bond hardness

It is important to understand the hardness of any material for practical device applications. Generally, some of the elastic properties such as Bulk modulus, shear modulus and Young's modulus give rough idea about hardness, but there is no direct one-to-one correspondence between hardness and Bulk/shear/Young's modulus [73,74]. With the help of Mulliken bond population, the Vickers hardness of non-metallic compounds can be calculated using a scheme formulated by F. Gao [74]. Since metallic bonding is delocalized and there is no defining relation between hardness and metallic bonding, this scheme [74] is not valid for crystals having partial metallic bonding. Considering a correction due to delocalized metallic bonding Gou *et al*. [75] reformulated the scheme to include compounds having partial metallic bond which can be expressed as follows:

$$H_v^\mu = 740(P^\mu - P^{\mu'})(v_b^\mu)^{-5/3} \tag{28}$$

where $P^\mu$ is the Mulliken population of the $\mu$-type bond, $P^{\mu'} = n_{free}/V$ is the metallic population (with $n_{free}$= number of free electrons), $v_b^\mu$ is the bond volume of $\mu$-type bond. The constant 740 is a proportionality coefficient obtained from the hardness of diamond. The geometric average of all bonds' hardness gives the total value of hardness for a complex multiband crystal, and can be expressed as [75,76]:

$$H_v = \left[\prod^\mu (H_v^\mu)^{n^\mu}\right]^{1/\Sigma n^\mu} \tag{29}$$

where $n^\mu$ is the number of $\mu$-type bond. The calculated theoretical bond hardness of SnTaS$_2$ is listed in Table 7 along with bond length, overlap population and bond volume. The positive and negative values of bond overlap population imply the presence of bonding-type and anti-bonding-type interactions between the atoms, respectively [32]. From Table 7, it is clear that both bonding-type and anti-bonding type interactions exist in SnTaS$_2$. Besides, a positive value of overlap population suggests presence of covalency in the bond. The level of covalency is more in the S-Ta bond than that in the S-Sn bond. The Vickers hardness of the compound of interest is 1.257 GPa. Therefore, SnTaS$_2$ is expected to be soft and easily machinable.

**Table 7.** The calculated Mulliken bond overlap population of $\mu$-type bond $P^\mu$, bond length $d^\mu$ (Å), metallic population $P^{\mu'}$, total number of bond $N^\mu$, bond volume $v_b^\mu$ (Å$^3$), hardness of $\mu$-type bond $H_v^\mu$ (GPa) and Vickers hardness, $H_v$ (GPa) of SnTaS$_2$.

| Compound | Bond | $P^\mu$ | $d^\mu$ | $P^{\mu'}$ | $N^\mu$ | $v_b^\mu$ | $H_v^\mu$ | $H_v$ |
|---|---|---|---|---|---|---|---|---|
| SnTaS$_2$ | S-Ta | 1.38 | 2.5035 | 0.0245 | 8 | 17.732 | 8.318 | 1.257 |
|  | S-Sn | -0.03 | 2.8019 | 0.0245 |  | 24.859 | -0.190 |  |

## 3.8 Optical Properties

The understanding of energy/frequency dependent optical parameters is essential to predict how a material will respond when electromagnetic radiation is incident upon it. In order to investigate possible optoelectronic applications of a compound, knowledge regarding the response of the compound to infrared, visible and ultraviolet spectra is important. To study the response of SnTaS$_2$ to incident photons, various frequency dependent optical parameters, namely, dielectric function $\varepsilon(\omega)$, refractive index $N(\omega)$, optical conductivity $\sigma(\omega)$, reflectivity $R(\omega)$, absorption coefficient $\alpha(\omega)$ and electron energy loss function $L(\omega)$ (where $\omega = 2\pi f$ is the angular frequency) are calculated and discussed in this section. The spectra for these parameters are depicted in Figure 8 for incident energy up to 25 eV and the electric field polarizations along [100], [010], and [001] crystallographic directions.

Figure 8(a) illustrates the optical absorption coefficient of SnTaS$_2$. Since the onset of absorption coefficient starts at zero photon energy, it is easy to predict that the material under study is metallic. This result is consistent with the band structure calculations. The onset of optical absorption is due to the free charge carriers within the conduction band. The absorption coefficient is quite high in the spectral region from ~3.5 eV to 19 eV, peaking around photon energy of 6.5 eV. It is seen that SnTaS$_2$ absorbs ultraviolet radiation very effectively. There is significant anisotropy in the optical absorption with respect to the direction of electric field polarization.

Figure 8(b) shows the real and imaginary parts of optical conductivity $\sigma(\omega)$. As photoconductivity starts from zero photon energy; SnTaS$_2$ has no optical band gap, which reconfirms its conducting nature. Real part of $\sigma(\omega)$ becomes almost zero at ~20.5 eV in agreement with the position of the loss peak. Once again, significant optical anisotropy is found in the optical conductivity spectra.

The real and imaginary parts of dielectric function $\varepsilon(\omega)$ is depicted in Figure 8(c). The real part of the dielectric constant is related to the electrical polarization of the material, while the imaginary part is linked with dielectric loss. The real part crosses zero at ~20.5 eV and after that it becomes unity. On the other hand, the imaginary part approaches to almost zero after 20.5 eV. Therefore, the plasma frequency of the compound of interest should be 20.5 eV above which it becomes transparent to incident electromagnetic radiation. Overall, the dielectric function exhibits metallic character consistent with band structure and electronic density of states features.

The frequency dependent loss function $L(\omega)$ of SnTaS$_2$ is displayed in Figure 8(d). The loss peak is found at ~20.5 eV. This peak marks the characteristic plasmon energy for the corresponding material. The plasma oscillations due to collective motions of the charge carriers are induced at this particular energy. It is interesting to note that the plasma energy coincides with the sharp falls in the absorption coefficient and reflectivity.

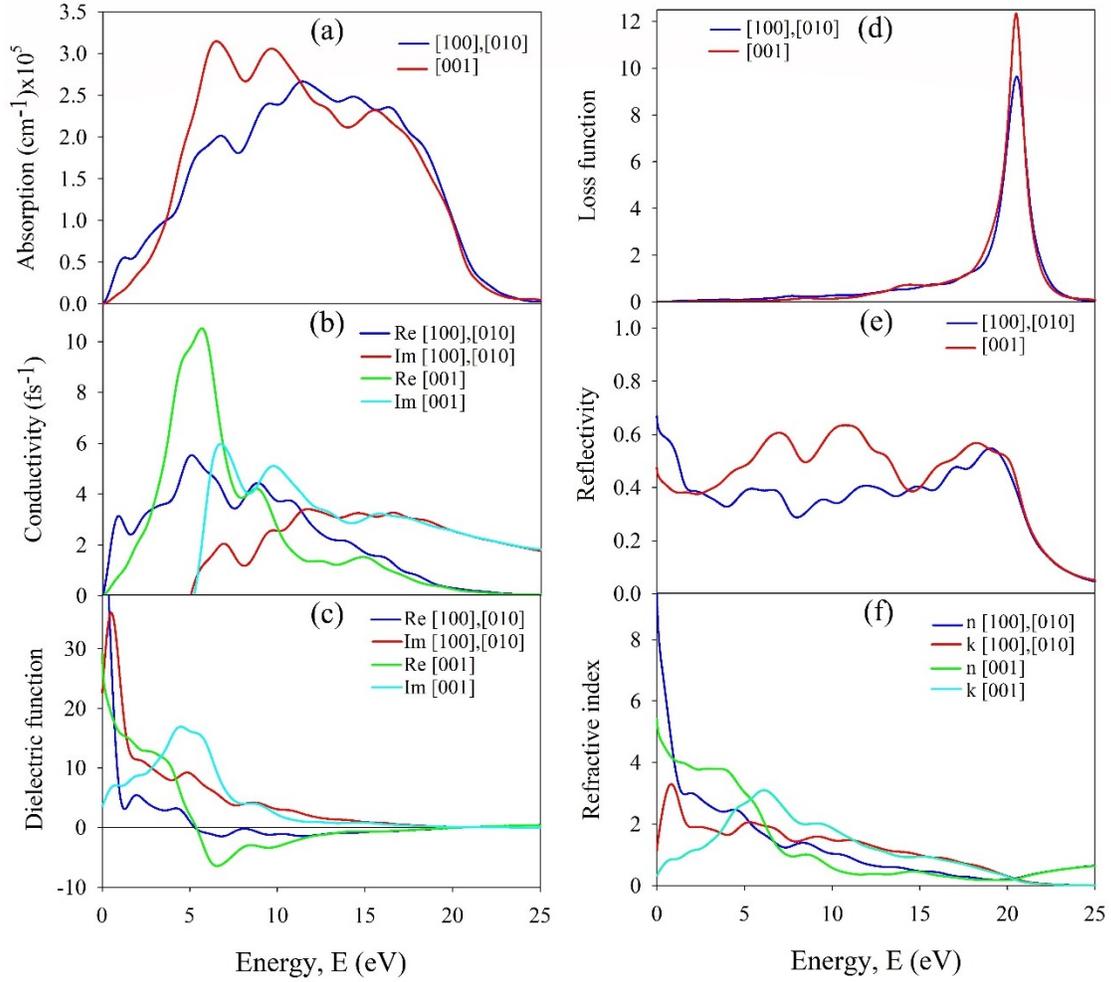

**Figure 8.** The frequency dependent (a) absorption coefficient (b) optical conductivity (c) dielectric function (d) loss function (e) reflectivity and (f) refractive index of SnTaS$_2$ with electric field polarization vectors along [100], [010], and [001] directions.

Figure 8(e) shows the reflectivity profile of SnTaS$_2$. The value of reflectivity for [001] direction is greater than [100] direction. For [001] electric field polarization, $R(\omega)$ remains 50% in the mid ultraviolet region (6 – 13 eV) and never goes below to ~ 40% from 0 to 20.5 eV. Moreover, in the both directions of polarization, reflectivity shows almost nonselective behavior over this broad energy range. Therefore, this material has wide band high reflectivity and can be employed as a reflector to reduce solar heating.

The frequency dependent refractive index is a complex parameter, expressed as $N(\omega) = n(\omega) + ik(\omega)$, where the imaginary part, $k(\omega)$, known as extinction coefficient is illustrated in Figure 8(f). The phase velocity of electromagnetic wave in the material is determined by the real part of refractive index, whereas the attenuation of electromagnetic radiation inside the material is measured by the imaginary part. The real part of the refractive index has high value at low photon energies including the visible region. Therefore, the compound under investigation has suitable optical characteristics for optoelectronic device applications. All the optical parameters of $SnTaS_2$ show anisotropy of varying degree with respect to the polarization direction of the electric field; anisotropy is least in loss function.

## 3.9 Acoustic velocities and its anisotropy

The velocity of sound in a material is an important parameter to study its mechanical properties, and thermal and electrical conductivities. Many crystals with high sound velocity, such as diamond and silicon carbide, are known as the best room temperature heat conductors. In this section, we have estimated the phase velocity of longitudinal and transverse modes of $SnTaS_2$. The transverse and longitudinal velocity of sound in a crystalline material can be determined with the help of Bulk modulus, $B$ and shear modulus, $G$, using the following equations [77]:

$$v_t = \sqrt{\frac{G}{\rho}} \quad \text{and} \quad v_l = \sqrt{\frac{B + 4G/3}{\rho}} \tag{30}$$

Here, $\rho$ is the mass-density of the solid. These equations predict that the sound velocities of a material are strongly dominated by its density and a material with zero shear modulus cannot support the transverse mode of sound propagation.

The average sound velocity $v_a$ in the crystal is determined, using the transverse and longitudinal sound velocities, from the following equation [77]:

$$v_a = \left[\frac{1}{3}\left(\frac{2}{v_t^3} + \frac{1}{v_l^3}\right)\right]^{-\frac{1}{3}} \tag{31}$$

The study of acoustic impedance of materials is useful which regulates the transfer of acoustic energy between two media. Over the years, the study of acoustic impedance of materials, having same or different impedance with the surrounding medium, has become a considerable tool for noise reduction, transducer design, aircraft engine manufacturing, industrial factories design and many underwater acoustic applications. The difference in acoustic impedance between a material and its surroundings determines the amount of acoustic energy transmitted and reflected at their interface when sound is transmitted from one medium (material) to another. When the impedance difference is large, most of the sound is reflected resulting in the loss of transmitted signal and echo generation. On the other hand, most of it is transmitted if the two impedances are about equal. The acoustic impedance of $SnTaS_2$ has been calculated using following equation [13,14]:

$$Z = \sqrt{\rho G} \tag{32}$$

where $G$ is the shear modulus and $\rho$ is the density of the material. The unit of acoustic impedance is the Rayl: 1 Rayl = $kgm^{-2}s^{-1}$ = 1 $Nsm^3$. A material with high density and high modulus of rigidity possesses high acoustic impedance.

The physics of musical instrument uses several important design parameters to describe and characterize its behavior. The intensity of sound radiation is one of those. This is a basic parameter for the designing of sound board and loudspeaker. The intensity, $I$, of acoustic radiation of a material is related to its shear modulus and density as [13,14,78]:

$$I \approx \sqrt{G/\rho^3} \tag{33}$$

where, $\sqrt{G/\rho^3}$ is defined as the *radiation factor*. The *radiation factor* is generally used for the selection of materials for sound board design. The calculated values of sound velocities, acoustic impedance and radiation factor for $SnTaS_2$ are enlisted in Table 8.

**Table 8.** Density $\rho$ (g/cm$^3$), transverse velocity $v_t$ (ms$^{-1}$), longitudinal velocity $v_l$ (ms$^{-1}$), average elastic wave velocity $v_a$ (ms$^{-1}$), acoustic impedance $Z$ (Rayl), and radiation factor $\sqrt{G/\rho^3}$ (m$^4$/kg.s) of $SnTaS_2$.

| Compound | $\rho$ | $v_t$ | $v_l$ | $v_a$ | $Z$ (×10$^6$) | $\sqrt{G/\rho^3}$ |
|---|---|---|---|---|---|---|
| $SnTaS_2$ | 7.09 | 2154.21 | 4144.68 | 2392.12 | 15.28 | 0.304 |

The velocity of sound (longitudinal and transverse) waves in a material is neither affected by its frequency nor the dimension of the material but its nature (material). For each atom in a system, only three modes of vibrations (one longitudinal and two transverse modes) are present. For anisotropic crystal, the pure longitudinal and transverse modes of waves are only possible along certain crystallographic directions, whereas the quasi-transverse or quasi longitudinal modes are present in all other directions. For crystals with hexagonal symmetry, the pure transverse and longitudinal modes can only exist for the high symmetry directions of type [100] and [001]. For hexagonal crystal, the acoustic velocities along these principle directions can be expressed as [79]:

*[100]*:

$$[100]v_l = \sqrt{(C_{11} - C_{12})/2\rho}; [010]v_{t1} = \sqrt{C_{11}/\rho}; [001]v_{t2} = \sqrt{C_{44}/\rho}$$

(34)

*[001]*:

$$[001]v_l = \sqrt{C_{33}/\rho}; [100]v_{t1} = [010]v_{t2} = \sqrt{C_{44}/\rho}$$

where $v_{t1}$ and $v_{t2}$ refer to the first transverse mode and the second transverse mode, respectively. Materials with lower density and higher elastic constants will have larger sound velocities in it. Directional sound velocities of SnTaS$_2$ are presented in Table 9.

**Table 9.** Anisotropic sound velocities (ms$^{-1}$) of SnTaS$_2$ along different crystallographic directions.

| Propagation directions | | SnTaS$_2$ |
|---|---|---|
| | $[100]v_l$ | 2790.06 |
| [100] | $[010]v_{t1}$ | 4654.69 |
| | $[001]v_{t2}$ | 1631.48 |
| | $[001]v_l$ | 5347.98 |
| [001] | $[100]v_{t1}$ | 1631.48 |
| | $[010]v_{t2}$ | 1631.48 |

Significant anisotropy in sound velocity is present in SnTaS$_2$ reflecting the layered structure of the compound and anisotropy in the bonding strengths along different crystallographic axes.

### 3.10 Thermophysical properties
### 3.10.1 Grüneisen parameter

The Grüneisen parameter, $\gamma$ is an important thermophysical parameter. This parameter gives an estimate of the anharmonic effects in a solid. A large value of $\gamma$ implies high level of anharmonicity. The Grüneisen parameter of SnTaS$_2$ compound is calculated from the Poisson's ratio using the following expression [80]:

$$\gamma = \frac{3(1+v)}{2(2-3v)}$$

(35)

Number of important physical processes, such as thermal conductivity, thermal expansion, absorption of acoustic waves and the temperature dependence of elastic properties, are linked to the Grüneisen parameter. The calculated value of $\gamma$ of SnTaS$_2$ is 1.87 (included in Table 10). This value is typical as most of the solids have Grüneisen parameters ~ 2.0.

### 3.10.2 Debye temperature

Debye temperature is the temperature at which the wavelength of phonon frequency roughly equals the unit cell length. The importance of Debye temperature comes from its relation to the thermophysical properties of solids. Number of thermophysical properties, for instance, lattice vibration, interatomic bonding, thermal conductivity, melting temperature, coefficient of thermal expansion, and phonon specific heat of solids are controlled by their Debye temperature. It also separates the high- and low-temperature regions. Materials with higher Debye temperature generally show stronger interatomic bonding strength, higher melting temperature, greater hardness, higher sound wave velocity and lower average atomic mass. At temperatures higher than $\Theta_D$, all modes of vibrations, discussed in Section 3.9, should possess an energy ~ $k_BT$. On the other hand, when the temperature is less than $\Theta_D$, the higher frequency modes are expected to be frozen [81]. Furthermore, Debye temperature is directly proportional to the superconducting transition temperature of phonon mediated superconductors [65,67]. There are two different methods widely used to determine Debye temperature, namely from specific heat measurements and using the elastic moduli. At low temperatures, Debye temperature calculated using elastic moduli is expected to be the same as that determined from specific heat measurements. We have calculated the Debye temperature of SnTaS$_2$ from the following equation [77,82]:

$$\Theta_D = \frac{h}{k_B}\left(\frac{3n}{4\pi V_0}\right)^{1/3} v_a \qquad (36)$$

where, $h$ is the Planck's constant, $k_B$ is the Boltzmann's constant, $V_0$ is the optimized volume of unit cell and $n$ is the number of atoms within the unit cell. The calculated value of the Debye temperature of SnTaS$_2$ is listed in Table 9. Debye temperature of SnTaS$_2$ is obtained as 199.63 K. The Debye temperature of SnTaS$_2$ is low compared to many other layered metallic ternaries [40,41,51,76,83-85] in accordance with its relatively soft nature.

Using the McMillan's formula [86] for superconducting transition temperature and the computed values of $\Theta_D$ and $\mu^*$, we have calculated the electron-phonon coupling constant of SnTaS$_2$. The estimated value of coupling constant is found to be 0.66, in exact agreement with the value found in Ref. [16].

### 3.10.3 Melting temperature

For understanding the temperature limit of materials to use, the study of melting temperature ($T_m$) is required. The thermal expansion, elastic constants and bonding energy of crystalline materials are correlated with their melting temperatures. Materials with higher melting temperatures have stronger atomic bonding and lower coefficient of thermal expansion [14]. Information regarding melting temperature of materials also helps us to get an idea about the temperature at which the materials can be used continuously without oxidation, chemical change, and excessive distortion due to heating becoming a problem. The computed values of elastic constants of $SnTaS_2$ can be used to determine its melting temperature $T_m$, using the following empirical equation [87]:

$$T_m = 354K + (4.5K/GPa)\left(\frac{2C_{11} + C_{33}}{3}\right) \pm 300K \tag{37}$$

The estimated value of melting temperature of $SnTaS_2$ is listed in Table 10. The melting temperature of $SnTaS_2$ is 1119.20 ± 300 K. There is a large uncertainty of ± 300 K in the calculated melting temperature $T_m$ of $SnTaS_2$. In the absence of any experimental or theoretical estimate of $T_m$ of $SnTaS_2$, the value quoted herein should be treated as a rough measure awaiting verification.

### 3.10.4 Thermal expansion coefficient and heat capacity

In solids, the atoms vibrate about their mean positions; the higher the temperature, the greater the amplitude of these vibrations. Thermal expansion coefficient ($\alpha$) is an intrinsic thermophysical parameter of solids. A number of other physical properties, such as thermal conductivity, specific heat, temperature variation of the energy band gap and electron effective mass of solids are correlated with the thermal expansion coefficient. The thermal expansion coefficient of a solid can be evaluated from the following equation [14,78]:

$$\alpha = \frac{1.6 \times 10^{-3}}{G} \tag{38}$$

where, $G$ is the isothermal shear modulus (in GPa). Thermal expansion of solids is inversely related to their melting temperature: $\alpha \approx 0.02/T_m$ [78,88]. The thermal expansion coefficient of $SnTaS_2$ is tabulated in Table 10.

Heat capacity is another crucial thermodynamic parameter of a material. Materials with higher heat capacity possess higher thermal conductivity and lower thermal diffusivity. The heat capacity per unit volume ($\rho C_P$) of a material defines the change in thermal energy per unit volume in a material per degree Kelvin change in temperature. The heat capacity of a material per unit volume in the high temperature limit can be calculated from [13,14,78]:

$$\rho C_P = \frac{3k_B}{\Omega} \tag{39}$$

where, $N = 1/\Omega$ is the number of atoms per unit volume.

The dominant phonon wavelength of a compound, $\lambda_{dom}$, defines the wavelength at which the phonon distribution is peaked. The wavelength of the dominant phonon for SnTaS$_2$ at 300 K is estimated by using following relationship [56,88]:

$$\lambda_{dom} = \frac{12.566 v_a}{T} \times 10^{-12} \tag{40}$$

where, $v_a$ is the average sound velocity in ms$^{-1}$, $T$ is the temperature in degree Kelvin. The obtained value of $\lambda_{dom}$ (in meter) of SnTaS$_2$ is listed in Table 10.

**Table 10.** The Debye temperature $\Theta_D$ (K), thermal expansion coefficient α (K$^{-1}$), wavelength of the dominant phonon mode at 300 K $\lambda_{dom}$ (m), melting temperature $T_m$ (K), heat capacity per unit volume $\rho C_P$ (J/m$^3$.K), and Grüneisen parameter $\gamma$ of SnTaS$_2$.

| Compound | $\Theta_D$ | $T_m$ | α (×10$^{-5}$) | $\lambda_{dom}$ (× 10$^{-12}$) | $\rho C_P$ (× 10$^6$) | $\gamma$ | Ref. |
|---|---|---|---|---|---|---|---|
| SnTaS$_2$ | 199.63 | 1119.20 | 4.86 | 100.20 | 1.95 | 1.87 | This work |
|  | 154.4 | - | - | - | - | - | [16] |

### *3.10.5 Minimum thermal conductivity and its anisotropy*
Information on the minimum thermal conductivity ($k_{min}$) of a material is required to control its high temperature applications. It is defined as the thermal conductivity of a material, which reaches to its minimum value at enough high temperature. Surprisingly, the minimum thermal conductivity of a material does not depend on the presence of defects, namely dislocations, individual vacancies and long-range strain fields associated with impurity inclusions. This is mainly because at high temperatures the phonon wavelength becomes comparable to interatomic spacing and perturbations distributed over longer length scales have negligible effect on thermal transport. The minimum thermal conductivity of a material is related to their sound velocity and Debye temperature. At high temperature, Clarke's formula has been used for calculating the minimum thermal conductivity $k_{min}$ of materials [88]:

$$k_{min} = k_B v_a (V_{atomic})^{-2/3} \tag{41}$$

where, $k_B$ is the Boltzmann constant, $v_a$ is the average sound velocity and $V_{atomic}$ is the average volume per atom [88]. The calculated value of minimum thermal conductivity for SnTaS$_2$ is listed in Table 11.

The transmission of heat through solids mainly takes place via the movement of free electrons in metals and thermal vibrations (phonons) and radiation (if they are transparent). Heat transmitted by thermal vibrations involves the propagation of elastic waves (or sound wave). An acoustically anisotropic material possesses anisotropy in minimum thermal conductivity. The anisotropy in minimum thermal conductivity of a compound can be estimated from Cahill's equation [89] given by:

$$k_{min} = \frac{k_B}{2.48} n^{2/3}(v_l + v_{t1} + v_{t2}) \tag{42}$$

and $$n = N/V$$

where, $k_B$ is the Boltzmann constant, $n$ is the number of atoms per unit volume and $N$ is total number of atoms in the cell with volume $V$. The equation explicitly tells that anisotropy in minimum thermal conductivity is controlled by the longitudinal and transverse sound velocities along different crystallographic directions. The minimum thermal conductivity of $SnTaS_2$ along [100] and [001] directions are presented in Table 11.

For the sake of completeness, we have also calculated minimum thermal conductivity of $SnTaS_2$ using both Cahill's and Clarke's models [56,89,90]. The calculated values are listed in Table 11. The $k_{min}$ of $SnTaS_2$ using Cahill's and Clarke's models are 0.62 and 0.44, respectively. It has been also observed from our previous studies that Clarke's model predicts lower minimum thermal conductivity compared to that predicted by Cahill's model [56,90].

**Table 11.** The number of atoms per mole of the compound $n$ (m$^{-3}$) and minimum thermal conductivity (W/m.K) of $SnTaS_2$ along different directions evaluated by Cahill's and Clarke's method.

| Compound | $n$ (10$^{28}$) | [100]$k_{calc.}^{Min}$ | [001]$k_{calc.}^{Min}$ | $k_{min}$ | |
|---|---|---|---|---|---|
| | | | | Cahill | Clarke |
| $SnTaS_2$ | 4.70 | 0.66 | 0.62 | 0.62 | 0.44 |

The thermal conductivity of $SnTaS_2$ is low compared to many other metallic ternaries [91]; a fact consistent with its relatively low Debye temperature [40,41,51,76,83-85].

## 4. Discussion and Conclusions

Large number of hitherto unexplored elastic, bonding, phonon, electronic, optical and thermophysical properties of recently discovered $SnTaS_2$ layered semimetal exhibiting type-II low-$T_c$ superconductivity have been investigated in detail via first-principles method in this work. $SnTaS_2$ is found to be mechanically and dynamically stable. The compound under study is highly ductile, machinable and possesses high level of dry lubricity. The material is soft. The mechanical and elastic features reflect the underlying bonding characters in SnTaS$_2$ which is dominated by ionic contribution with some additional contributions originating from covalent and metallic channels. SnTaS$_2$ is elastically anisotropic with a highly layered crystal structure.

Electronic band structure features reveal a metallic character with anisotropy in the energy dispersion. The location of the Fermi level separates dip and peak in the TDOS nearby. This, together with high compressibility of the compound, implies that application of pressure can induce large change in electronic and optical properties including superconductivity. The layered character of the compound also indicates that band structure tailoring with intercalation with suitable atoms is a possibility in SnTaS$_2$. This can open up opportunities to tune optoelectronic and superconducting state properties of SnTaS$_2$. The repulsive Coulomb pseudopotential of SnTaS$_2$ is high. The Fermi surface has both electron- and hole-like sheets.

The optical parameters spectra of SnTaS$_2$ exhibit several interesting features. The reflectivity spectrum is largely nonselective over a wide spectral range of photon energy and the material can be used as a solar reflector. The absorption coefficient is high in the mid ultraviolet region and the low energy refractive index is high. Overall, the optical parameters spectra conform well to the band structure and electronic density of states results. The material possesses optical anisotropy.

The Debye temperature and thermal conductivity of SnTaS$_2$ are low in agreement with the phonon structure, bonding characteristics and elastic properties calculations. The electron-phonon coupling constant has been calculated and shows excellent agreement with that calculated in a previous study [16]. Calculated value of the coupling constant discloses that SnTaS$_2$ is a weakly coupled [65,92] low transition temperature superconductor.

Most of the parameters disclosed in Tables 2 – 11 are novel. No theoretical or experimental values exist in the literature for comparison. Therefore, the computed parameters should be treated as predictions and should be used as reference values for future studies.

We hope that the study presented in this paper will inspire researchers to investigate the physical properties of SnTaS$_2$ in further details, both experimentally and theoretically, in future.

**Acknowledgements**
S. H. N. and R. S. I. acknowledge the research grant (1151/5/52/RU/Science-07/19-20) from the Faculty of Science, University of Rajshahi, Bangladesh, which partly supported this work.

**Data availability**
The data sets generated and/or analyzed in this study are available from the corresponding author on reasonable request.

**Author Contributions**

M.I.N. performed the theoretical calculations, contributed to the analysis and contributed to draft manuscript. M.M. performed theoretical calculations, contributed to the analysis and contributed to draft manuscript. A.T. contributed to the analysis and contributed to draft manuscript. R.S.I. supervised the project and contributed in manuscript writing. S.H.N. designed and supervised the project, analyzed the results and finalized the manuscript. All the authors reviewed the manuscript.

**Additional Information**
**Competing Interests**
The authors declare no competing interests.